\documentclass[preprint,11pt,3p]{elsarticle}
\usepackage{amssymb}
\usepackage[usenames]{xcolor}
\usepackage{xspace}
\usepackage{subfig}
\usepackage[dvips]{epsfig}
\usepackage{epstopdf}
\usepackage{array}
\usepackage{framed}
\usepackage{comment}
\usepackage[normalem]{ulem}

\usepackage{moreverb}

\newcommand\BibTeX{{\rmfamily B\kern-.05em \textsc{i\kern-.025em b}\kern-.08em
T\kern-.1667em\lower.7ex\hbox{E}\kern-.125emX}}
\journal{International Journal for Numerical Methods in Biomedical Engineering}

\begin{document}
\begin{frontmatter}

\title{Non-uniform force allocation\\ for area preservation in spring network models\footnote{This work was supported by the Slovak Research and Development Agency, contract No. APVV-0441-11.}}

\author{Ivan Cimr\'ak\footnote{The work of I. Cimr\'ak was supported by the the Marie-Curie grant No. PCIG10-GA-2011-303580.}}
\author{Iveta Jan\v cigov\'a}
\address{Faculty of Management Science and Informatics,\\ University of \v Zilina, Slovakia\\e-mail: iveta.jancigova@fri.uniza.sk}

\begin{abstract}
In modelling of elastic objects in a flow such as red blood cells, white blood cells, or tumour cells, several elastic moduli are involved. One of them is the area conservation modulus. In this paper, we focus on spring network models and we introduce a new way of modeling the area preservation modulus. We take into account the current shape of the individual triangles and find the proportional allocation of area conservation forces, which would for individual triangles preserve their shapes. The analysis shows that this approach tends to regularize the triangulation. We demonstrate this effect on individual triangles as well as on the complete triangulations.
\end{abstract}


\end{frontmatter}

%\vspace{-6pt}
%%%%%%%%%%%%%%%%%%%%%%%%%%%%%%%%%%%%%%%
% 
%        section
%
%%%%%%%%%%%%%%%%%%%%%%%%%%%%%%%%%%%%%%%

\section{Introduction}
The elasticity of red blood cell (RBC) plays an important role in blood flow - both in the circulatory system where the cell can deform to such extent that it is able to pass through capillaries with diameter smaller than its size~\cite{Gaehtgens1980} and in microfluidic devices where again the RBC deformations might enable the cell to pass through structures meant to capture some other - not so elastic - cells from blood~\cite{Xu2013b}.\\
It is widely accepted that the RBC in its relaxed (zero-stress) state is a biconcave discoid. While few studies suggest that the stress-free state of an RBC might approach a sphere \cite{Peng2014}, our model is based on the biconcave discoid hypothesis. The RBC has an elastic outer membrane, which is filled with viscous inner fluid and immersed organelles. The membrane consists of a lipid bilayer with high resistance to areal change and an underlying spectrin network. These two are connected by transmembrane proteins~\cite{Mohandas2008}. The membrane is primarily responsible for the elastic behavior of the cell and thus it is the focus of our modeling.\\
This study investigates the modulus that captures area conservation of the membrane.\\
The models of red blood cells are often based on spring networks \cite{Fedosov2010b,Dupin2007,Dao2006}. In these models, the area conservation modulus is implemented in various ways. In our earlier investigations \cite{Cimrak2012,Cimrak2013a,Jancigova2014a}, we have used an RBC model, in which the implementation of area preservation modulus was inspired by \cite{Dupin2007}. This implementation does not fulfil the force-free and torque-free conditions and may worsen the mesh regularity under certain conditions. This prompted us to look for a better approach to local area conservation.\\
The work is organized as follows: In Section \ref{sec:forces}, we summarize the model we have originally used, which - while functional - may introduce irregularities of the underlying spring network when the object undergoes large deformations. In Section \ref{sec:mesh_regularity}, we review the criteria for mesh regularity needed to evaluate the new local area preservation forces that we propose in Section \ref{sec:definition}. In the next Section \ref{sec:force_free} we prove that this implementation is force-free and torque-free both globally and locally. Further, we compare it to the previous implementation both analytically (Section \ref{sec:analytical}) and using simulations (Section \ref{sec:ellipsoids}). We show that the new implementation guarantees better properties of the surface triangulation as the object undergoes deformations.\\    

%%%%%%%%%%%%%%%%%%%%%%%%%%%%%%%%%%%%%%%
% 
%        section
%
%%%%%%%%%%%%%%%%%%%%%%%%%%%%%%%%%%%%%%%

\section{Model}\label{sec:forces}
Our model comes from \cite{Dupin2007} and it comprises two main parts - the fluid and the elastic object. For the fluid, e.g. blood plasma, we use the lattice-Boltzmann method (as described in \cite{Dunweg2009}). For elastic objects, e.g. the red blood cells, we use a spring network model described in the next section, in which the spring network forms a surface triangulation of the object. These two models are connected using the immersed boundary method \cite{Feng2004}, in which the movement of immersed boundary points is governed by Newton's equation of motion.\\

Our primary objects of interest are red blood cells and therefore the elastic properties that we need captured in the model are resistance of the membrane to surface dilation, elastic resistance to bending and stretching of the membrane and total volume conservation. Discussion of continuum approach to 2D elasticity may be found e.g. in \cite{Pozrikidis2003}, but in the model described in this paper, the stretching coefficient does not directly correspond to the Young's modulus \cite{Cimrak2013a} or local/global area coefficients to the area dilatation modulus. We define the five elastic forces to conserve the respective quantities (lengths, angles, areas, volume) and calibrate their coefficients so that the overall behavior matches the cell behavior seen in biological experiments. Here we provide a description of the elastic forces that we use (more details can be found in \cite{Cimrak2012,Tothova2014b}). Stretching force: 
\begin{eqnarray}\label{eq:stretching_equation}
\mathbf{F}_s(A) =k_s \kappa (\lambda_{AB}) \Delta L_{AB}\mathbf{n}_{AB}\\
\kappa(\lambda_{AB}) = \frac{\lambda_{AB}^{0.5}+\lambda_{AB}^{-2.5}}{\lambda_{AB}+\lambda_{AB}^{-3}}
\end{eqnarray}
where \(k_s\) is the stretching coefficient, $\lambda_{AB}=L_{AB}/L^0_{AB}$, $\kappa(\lambda_{AB})$ represents the neo-Hookian nonlinearity of the stretching force (taken from \cite{Dupin2007}). $L^0_{AB}$ is the relaxed length of the edge $AB$, $\Delta L_{AB}=L_{AB}-L^0_{AB}$ is the prolongation of this edge and $\mathbf{n}_{AB}$ is the unit vector pointing from $A$ to $B$. Bending force:
\begin{equation}\label{eq:bending_equation}
\mathbf{F}_b(A) =k_b \frac{\Delta \theta}{\theta} \mathbf{n}_{ABC}
\end{equation} 
where \(k_b\) is the bending coefficient, \(\theta\) is the resting angle between two triangles that have common edge \(BC\), \(\Delta \theta\) is the deviation from this angle and \(\mathbf{n}_{ABC}\) is the unit normal vector to the triangle \(ABC\). The forces on the vertices $B$ and $C$ from the common edge have the opposite direction.
Local area force:
\begin{equation}\label{eq:local_area_equation}
\mathbf{F}_{al}(A) =k_{al} \frac{\Delta S_{ABC}}{\sqrt {S}} \mathbf{w}_{A}
\end{equation} 
where \(k_{al}\) is the local area coefficient, \(\Delta S_{ABC}\) is the deviation from the area in resting state, $S$ is the current area of the triangle, and \(\mathbf{w}_{A}\) is the unit vector pointing from the vertex \(A\) to the centroid of the triangle \(ABC\). (Analogous forces are assigned to vertices \(B\) and \(C\)). Global area force:
\begin{equation}\label{eq:global_area_equation}
\mathbf{F}_{ag}(A) =k_{ag} \frac{\Delta S}{S} \mathbf{w}_{A}
\end{equation} 
where \(k_{ag}\) is the global area coefficient, \(S\) is the relaxed area of the whole object, \(\Delta S\) is the deviation from this area and \(\mathbf{w}_{A}\) is again the unit vector pointing from the vertex \(A\) to the centroid of the triangle \(ABC\). Volume force:
\begin{equation}\label{eq:volume_equation}
\mathbf{F}_v(A) =-k_v \frac{\Delta V}{V} S_{ABC}\mathbf{n}_{ABC}
\end{equation} 
where \(k_v\) is the volume coefficient, \(S_{ABC}\) is the area of triangle \(ABC\), \(V\) is the volume of the whole object, \(\Delta V\) is the deviation from this relaxed volume and \(\mathbf{n}_{ABC}\) is the unit normal vector to the plane \(ABC\). 

A discussion of scalability of forces can be found in \cite{Jancigova2014a} and our analysis of influence of mesh density on the elastic coefficients is summarised in \cite{Tothova2015a}.  

Numerous authors \cite{Fedosov2010a,Odenthal2013,Nakamura2013, Ujihara2010} use different approach for definition of elastic forces. They start from the energy of the spring system. For the local area preservation modulus, this gives the following contributions for each triangle to the overall energy 
\begin{equation}
E_{al}=\frac{k_{al}}{2}\Delta S^2/S.
\end{equation}
The corresponding forces are obtained by differentiation of the energy with respect to the position. In this particular example, the force has the direction from the vertex of the triangle perpendicular to the opposite edge of the triangle \cite{Fedosov2010}. There is however one significant disadvantage of this approach. This force applied to the vertices of an obtuse triangle results in enlarging the obtuse angle. This makes the triangle more degenerate and the regularity of the triangulation is decreased, see Section \ref{sec:mesh_regularity} for extended discussion on the mesh regularity. 

On the other hand, the energy approach has its merit because it allows a straightforward comparison of acting elastic moduli and investigation of minimum energy states. There has been a simplified alternative proposed in \cite{Tothova2014a} that can be used instead of energy. however, one of the motivations for the new definition of area conservation forces was precisely the possibility to compute the actual elastic energy of other moduli besides stretching.

%%%%%%%%%%%%%%%%%%%%%%%%%%%%%%%%%%%%%%%
% 
%        section
%
%%%%%%%%%%%%%%%%%%%%%%%%%%%%%%%%%%%%%%%

\section{Mesh regularity and its quality}\label{sec:mesh_regularity}
Surface triangulations are frequently used in a large variety of scientific applications. Usually they are required (e.g. by the numerical method or by the physical characteristics of the triangulated surface) to meet specific properties and be reasonably faithful approximations of the true surface they represent. 

For the investigation of different implementations of local area conservation law we need to asses the quality of the mesh. A natural requirement is whether the mesh resembles the actual shape of the object. This however, can only be evaluated when we know the explicit expression for the object's shape. In such cases, we can evaluate an $L^2$ distance of mesh points from the object's surface.

In most cases however, we do not know the actual shape of the object, since it is to be approximated. Therefore, we can only measure the quality of the mesh using general principles that are shape independent. These principles are based on numerical aspects and are usually formulated using two criteria \cite{Chew1993}:\\
- each triangle must be well sized\\
- each triangle must be well shaped

The first statement is linked to the refinement techniques. In finite element methods, the mesh is often refined in order to better capture the high gradient of the approximated function or locally higher curvature of approximated surface. In our case, we do not use refinement since the surfaces are considered to have quite regular distribution of curvature. Our assumption is that the meshes in relaxed state have triangles of roughly the same sizes.

The second statement can be considered from several points of view. Some authors require that each angle in a triangle must be larger than $30^\circ$ \cite{Chew1993}. Other approaches look at distribution of length of edges (they should be roughly the same size), at distribution of angle sizes of triangles (they should be close to $60^\circ$), at aspect ratio between the radii of inscribed and circumscribed circles \cite{Frey1999}.

From computational point of view, it is reasonable to require that a mesh consists of triangles that are not degenerated. By degenerated we mean that two angles of a triangle have significantly different sizes. Or, in other words, triangles are non-degenerated when they are close to equilateral.

\def\Qt{Q_{T}}
In \cite{Frey1999}  the authors use the following expression that defines the shape quality of triangle $T$:
\begin{equation}\label{eq:shape_quality}
\Qt =\alpha \frac{\rho_{T}}{h_{max}}
\end{equation} 
where \(\alpha = 2\sqrt{3}\) is a normalization coefficient, \(\rho_{T}\) is the radius of inscribed circle and \(h_{max}\) represents the length of the longest edge. The shape quality varies in the interval (0,1], where values close to 0 are associated with an obtuse triangle with largest angle close to $\pi$. On the other hand, almost equilateral triangles have values of $\Qt$ close to one.

%%%%%%%%%%%%%%%%%%%%%%%%%%%%%%%%%%%%%%%
% 
%        section
%
%%%%%%%%%%%%%%%%%%%%%%%%%%%%%%%%%%%%%%%

\section{Proportional local area force}\label{sec:definition}

The main idea is to base the local area force definition on the difference \(\Delta S=S-S_0\) as it is in (\ref{eq:local_area_equation}), but instead of equal distribution of partial forces among the three triangle nodes, to find suitable proportions for the distribution. There are two criteria to keep in mind:\\
- the units of the local area coefficient \(k_{al}\) should be \(\left[\frac{N}{m}\right]\),\\
- the energy computed using this new local area force should be shape independent, i.e. the energy of two triangles that have the same area but different shape should be the same.

We derive the proportional force in such a way that it does not influence shape - the resulting triangle (in the ideal case, when there are no other forces present), is congruent to the original one, since this force is meant to conserve only the triangle area. And secondly, we want to use the same principle for computing the local area elastic energy, as one would use for stretching force (\ref{eq:stretching_equation}):
\begin{equation}
E_s= \int_{0}^{d} F_s(x)dx,
\end{equation}
which for linear stretching force \(F_s=k_s\Delta L\) gives \(E_s=\frac{1}{2}k_s (\Delta L)^2\).

\begin{figure}
 \includegraphics[width=0.45\textwidth]{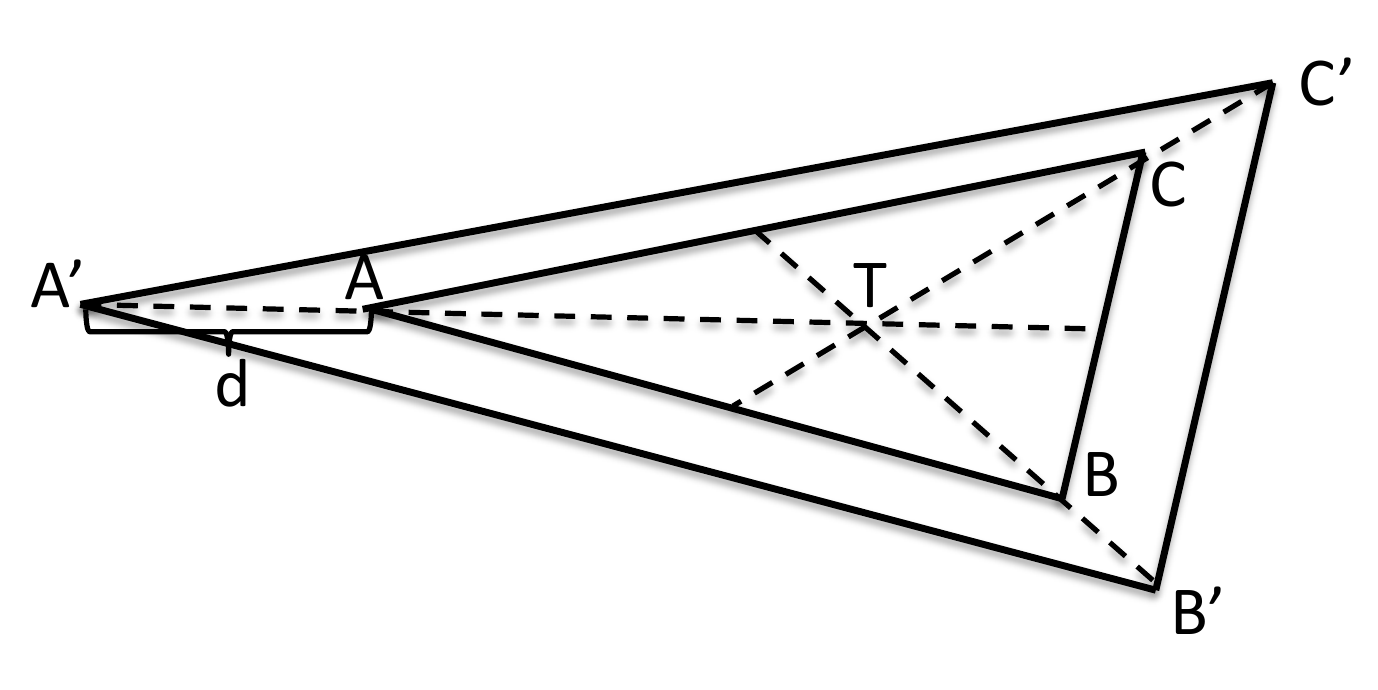}\hfill
 \includegraphics[width=0.45\textwidth]{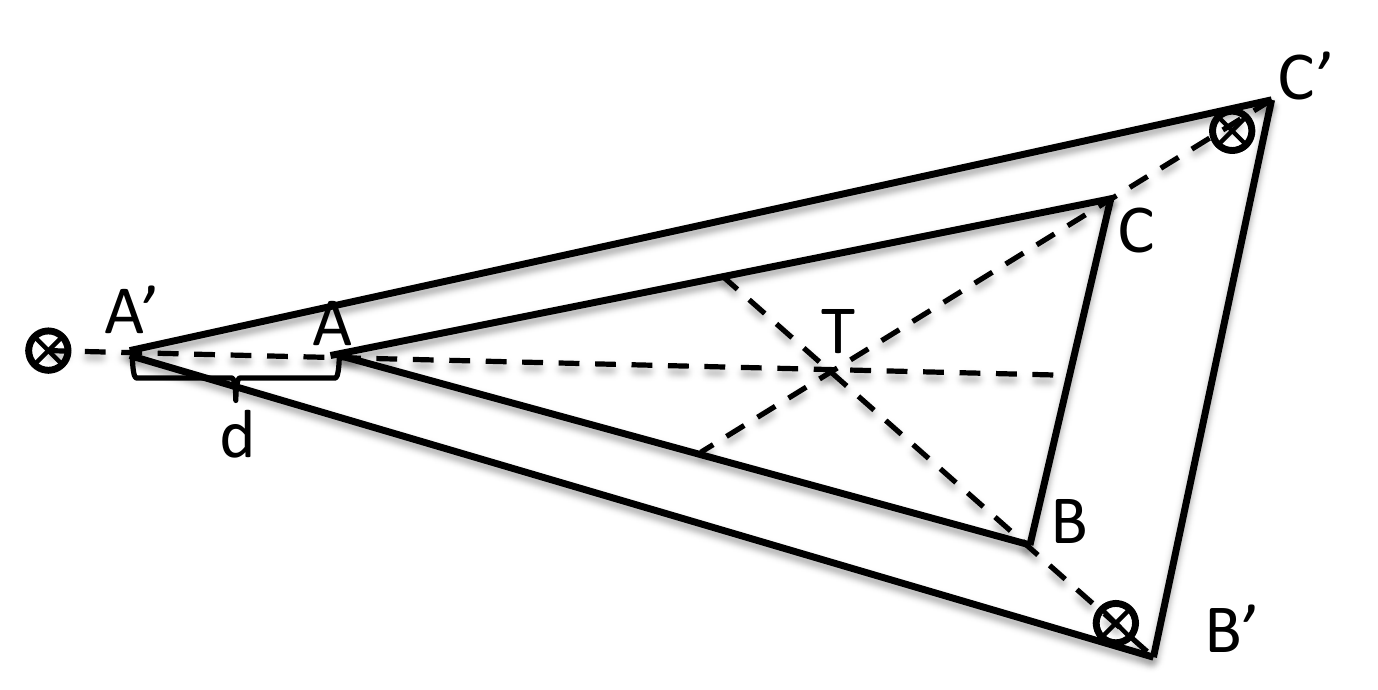}
 \caption{Left: Triangle \(ABC\) stretched to triangle \(A'B'C'\) using proportional implementation (\ref{eq:new_local_area_equation}). Depicted with medians and centroid $T$. Right: Triangle \(ABC\) stretched to triangle \(A'B'C'\) with the uniform implementation using expression (\ref{eq:local_area_equation}). The positions of stretched vertices using proportional implementation are indicated with three crossed circles.} \label{fig:triangle_sim}
\end{figure}

Suppose we start with triangle \(ABC\) and stretch it to triangle \(A'B'C'\) as shown in Figure \ref{fig:triangle_sim} left. We denote by $T$ the centroid (intersection of medians) and by \(d\) the distance \(|AA'|\). In order to simplify the notation, we also denote \(t_A=|AT|, t_B=|BT|, t_C=|CT|, t_A'=|A'T|, t_B'=|B'T|, t_C'=|C'T|\) and the similarity coefficient by \(t\). We can then write the following: \(d=t.t_A\) and \(|A'T|=(1+t)|AT|=(1+t)t_A, |B'T|=(1+t)t_B, |C'T|=(1+t)t_C\).

For areas we have \(S_{A'B'C'}=(1+t)^2S_{ABC}\) and consequently
\begin{equation}
\Delta S=S_{A'B'C'}-S_{ABC}=(2t+t^2)S_{ABC}.
\end{equation}
The magnitude of the local area force is then
\begin{equation}
F_{al}=k_{al}(2t+t^2)S_{ABC}.
\end{equation}

This force is applied at the three triangle nodes and in each case, its direction is towards the centroid \(T\). However, instead of applying the same force at the three vertices, we introduce proportionality coefficients \(p_i, i=A,B,C, F_i=F_{al}(i)=p_iF_{al}\). 

In order to determine the values of \(p_i\), we integrate using substitution \(y=\frac{x}{t_A}\) 
\begin{eqnarray}\nonumber
E_A  & = & \int_0^d F_A(x)dx=\int_0^t F_A(y) t_A dy  =  \int_0^t k_{al}(2y+y^2)S_{ABC}p_A t_A dy\nonumber\\
     & = & k_{al}t_Ap_AS_{ABC}(t^2+\frac{t^3}{3}) = k_{al} t_A' p_A S_{ABC}\frac{t^2+t^3/3}{1+t}
\end{eqnarray}

The total local area energy of one triangle is then
\begin{equation}
E_{al}= E_A+E_B+E_C=k_{al}S_{ABC}\frac{t^2+t^3/3}{1+t}\underbrace{(t_A'p_A + t_B'p_B + t_C'p_C)}_{M}
\end{equation}
In order to meet the two criteria mentioned at the beginning of this section, the quantity \(M\) has to be dimensionless and shape independent, i.e. should not contain the quantities \(t_A', t_B', t_C'\). This is true for 
\begin{equation}
p_i=\frac{t_i'}{t_A'^2+t_B'^2+t_C'^2}, i=A,B,C
\end{equation}
The new proposed local area force definition is then 

\begin{equation}\label{eq:new_local_area_equation}
F_{al}(A')=\frac{t_A'}{t_A'^2+t_B'^2+t_C'^2}k_{al}\Delta S \quad = \quad \frac{t_A'^2}{t_A'^2+t_B'^2+t_C'^2}k_{al}\frac{\Delta S}{t_a'}
\end{equation}
and analogous for vertices $B'$ and $C'$. 

\begin{figure}
 \begin{center}
 \includegraphics[width=0.45\textwidth]{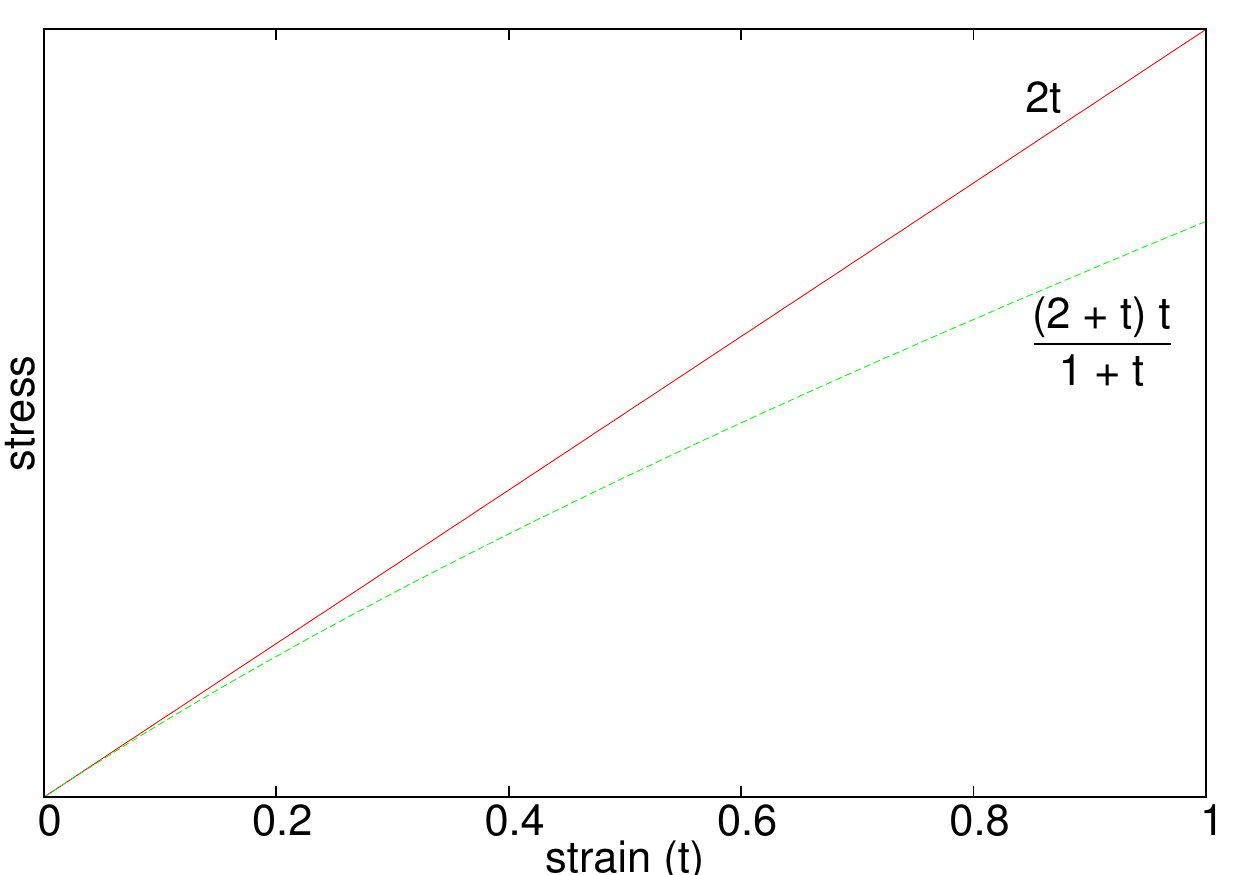}
 \caption{Strain-softening of the proportional implementation.} \label{fig:softening}
 \end{center}
\end{figure}

In blood cell modeling, it is interesting to determine whether the model exhibits strain-softening or strain-hardening behavior at large deformations~\cite{Chen2014}. 
We can analyze the equation (\ref{eq:new_local_area_equation}) using equilateral triangles. For non-equilateral triangles, the analysis would be analogous, since the proportional implementation generates congruent triangles with the ratio $t_a:t_b:t_c$ fixed. For equilateral triangles, we have 
\begin{equation}\label{eq:softening}
F_{al}(A') = \frac13k_{al}\frac{(1+t)^2 S_0 - S_0}{(1+t)t_a} = \frac13k_{al} \frac{S_0}{t_a}\frac{(2+t)t}{1+t}.
\end{equation}
The last fraction defines nonlinear relation between strain and stress. Linearization of this fraction in isotropic small deformation regime (for $t \to 0$) gives us function $2t.$ In Figure \ref{fig:softening}, we clearly see that in large deformation regime (for $t\in (0,1)$), the proportional implementation shows light strain-softening. 

%%%%%%%%%%%%%%%%%%%%%%%%%%%%%%%%%%%%%%%
% 
%        section
%
%%%%%%%%%%%%%%%%%%%%%%%%%%%%%%%%%%%%%%%

\section{Force-free and torque free conditions}\label{sec:force_free}
An important property that spring network models should satisfy are the force-free and torque-free conditions. The uniform implementation does not fulfil these conditions. To demonstrate this, we have designed the following computational experiment. We take a tetrahedron with vertices (0,0,0), (1,0,0), (0,1,0), (0,0,1) with only local area modulus (no other elastic forces are at play). We stretch the tetrahedron so that first, third and fourth vertices remain intact and the second is relocated to position (5,0,0). The tetrahedron starts shrinking due to the local area preservation forces. We track the position of its centroid. This experiment is run twice, once with the uniform formula (\ref{eq:local_area_equation}), and once with the proportional implementation (\ref{eq:new_local_area_equation}). Uniform local area forces cause the movement of centroid by 0.48 in $x$-direction, while the proportional local area keeps the centroid in place.

Next, we rigorously prove the force-free and torque-free conditions for proportional local area force both locally and globally. Local sum of forces on each triangle is given by 
\begin{eqnarray}
\mathbf{F}_{al}(A) + \mathbf{F}_{al}(B) + \mathbf{F}_{al}(C) &=& \frac{1}{t_A^2+t_B^2+t_C^2}k_{al}\Delta S\left(t_A \frac{\vec{AT}}{|AT|}+t_B \frac{\vec{BT}}{|BT|}+t_C \frac{\vec{CT}}{|CT|}\right)\nonumber \\ 
& = & \frac{1}{t_A^2+t_B^2+t_C^2}k_{al}\Delta S\left(\vec{AT} + \vec{BT} + \vec{CT} \right)\nonumber \\ 
& = & 0,
\end{eqnarray}
using the known fact that the sum of vectors pointing from vertices to the centroid of the triangle is zero. Consequently, the global force-free condition is fulfilled automatically.

The torque-free condition means that the object is not rotating around any point $X$ in space. First, we compute the local torque around centroid $T$ of the triangle 
\begin{equation}
\vec{TA} \times \mathbf{F}_{al}(A) + \vec{TB} \times \mathbf{F}_{al}(B) + \vec{TC} \times \mathbf{F}_{al}(C) = 0, 
\end{equation}
since the forces from individual vertices are co-linear with the lever-arm distance vector from the centroid. When computing the torque of the triangle around an arbitrary point $X$ in space we have 
\begin{eqnarray}
&\vec{XA} \times \mathbf{F}_{al}(A) + \vec{XB} \times \mathbf{F}_{al}(B) + \vec{XC} \times \mathbf{F}_{al}(C) = &  \nonumber \\
&(\vec{XT} + \vec{TA}) \times \mathbf{F}_{al}(A) + (\vec{XT}+\vec{TB}) \times \mathbf{F}_{al}(B) + (\vec{XT}+\vec{TC}) \times \mathbf{F}_{al}(C) = & \nonumber \\
&\vec{XT} \times (\mathbf{F}_{al}(A) + \mathbf{F}_{al}(B) +  \mathbf{F}_{al}(C)) + & \nonumber \\
&\vec{TA} \times \mathbf{F}_{al}(A) + \vec{TB} \times \mathbf{F}_{al}(B) + \vec{TC} \times \mathbf{F}_{al}(C) = 0,& 
\end{eqnarray}
since the force-free condition holds.

Global torque around the centroid $W$ of the object is computed as a sum over all triangles of torques associated to those triangles 
\begin{equation}
\sum_{\triangle_i \in triangles} \big(\vec{WA_i} \times \mathbf{F}_{al}(A_i) + \vec{WB_i} \times \mathbf{F}_{al}(B_i) + \vec{WC_i} \times \mathbf{F}_{al}(C_i)\big), 
\end{equation}
where $\triangle_i$ are the triangles belonging to the triangular mesh and $A_i, B_i, C_i$ are the vertices of triangle $\triangle_i$. Denote $T_i$ the centroid of $\triangle_i$ and split $\vec{WA_i} = \vec{WT_i} + \vec{T_iA_i}.$
The global torque can then be rewritten as
\begin{eqnarray}
\sum_{\triangle_i \in triangles}&&\vec{WT_i} \times \big(\mathbf{F}_{al}(A_i) +  \mathbf{F}_{al}(B_i) + \mathbf{F}_{al}(C_i)\big) + \nonumber\\
+ \sum_{\triangle_i \in triangles}&&\big(\vec{T_iA_i} \times \mathbf{F}_{al}(A_i) + \vec{T_iB_i} \times \mathbf{F}_{al}(B_i) + \vec{T_iC_i} \times \mathbf{F}_{al}(C_i) = 0\big).
\end{eqnarray}
The first term is zero because of the force-free condition and the second term is zero because it consists of cross products of co-linear vectors. 

Global torque around any point $X$ in space can be computed in the same way and yields the same result.

%%%%%%%%%%%%%%%%%%%%%%%%%%%%%%%%%%%%%%%
% 
%        section
%
%%%%%%%%%%%%%%%%%%%%%%%%%%%%%%%%%%%%%%%

\section{Analytical example}\label{sec:analytical}
We would like to compare the two implementations, the uniform one given by (\ref{eq:local_area_equation}) and the proportional one given by (\ref{eq:new_local_area_equation}). In Section \ref{sec:mesh_regularity}, we have concluded that the quality of triangular meshes depends on whether the individual mesh triangles are close to equilateral. Our analysis of uniform and proportional implementation of local area force is based on this observation. We use the measure given by (\ref{eq:shape_quality}). 

The benchmark test is designed in the following way. An equilateral triangle with all edge lengths equal to 1 is considered as an original triangle. This triangle is deformed to different initial triangles and we let each initial triangle relax under the influence of local area force. At the end of the relaxation process, we obtain a relaxed triangle. This relaxation process is run twice, once for the uniform implementation and once for the proportional implementation. The relaxed triangle does not have the same shape as  the original triangle because the local area force does not remember the original shape, only the area. After the triangle reaches its relaxed shape, we measure how well the relaxed triangle resembles the original equilateral shape using (\ref{eq:shape_quality}).

\begin{figure}
 \includegraphics[width=0.45\textwidth]{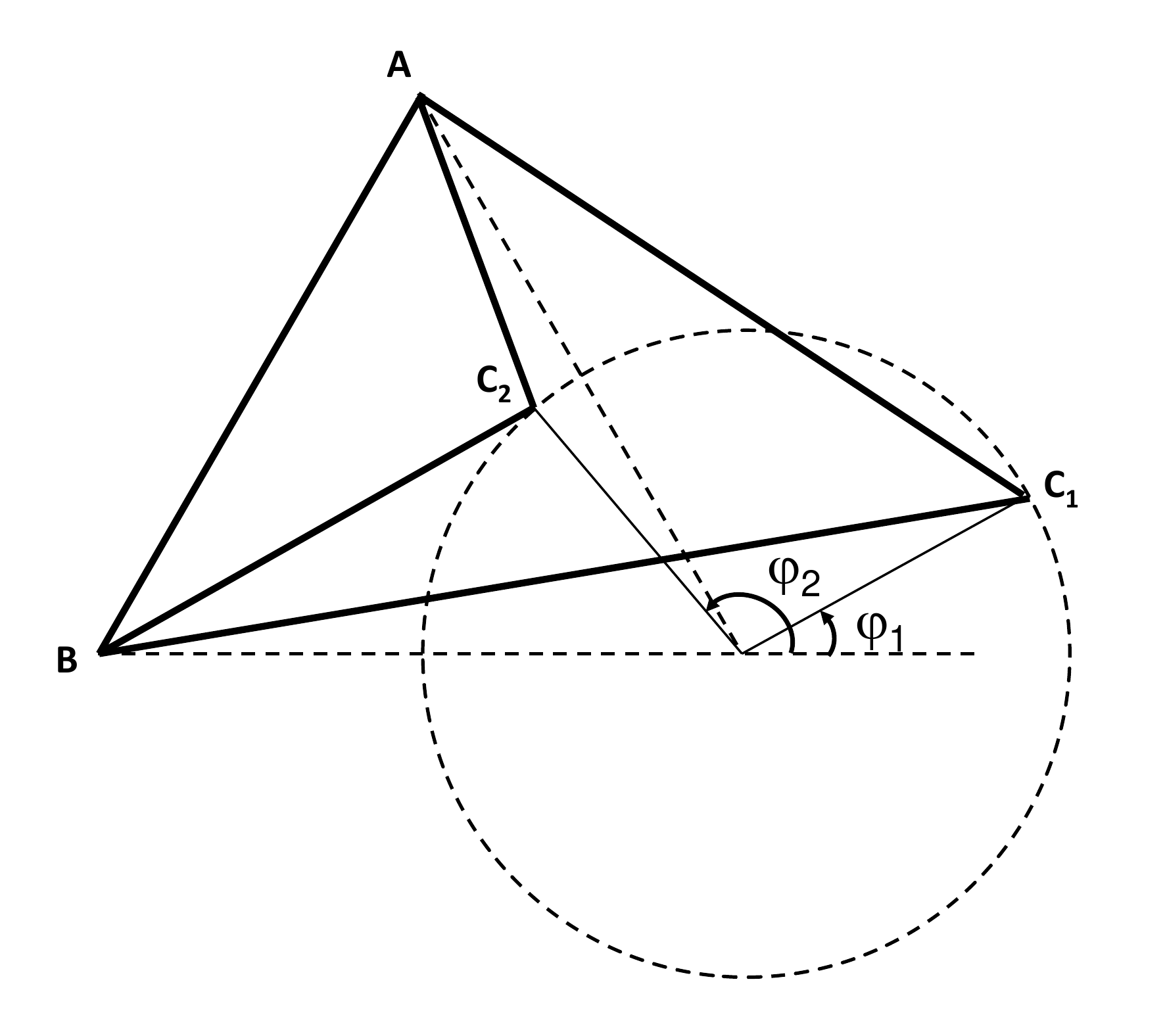}\hfill
 \includegraphics[width=0.45\textwidth]{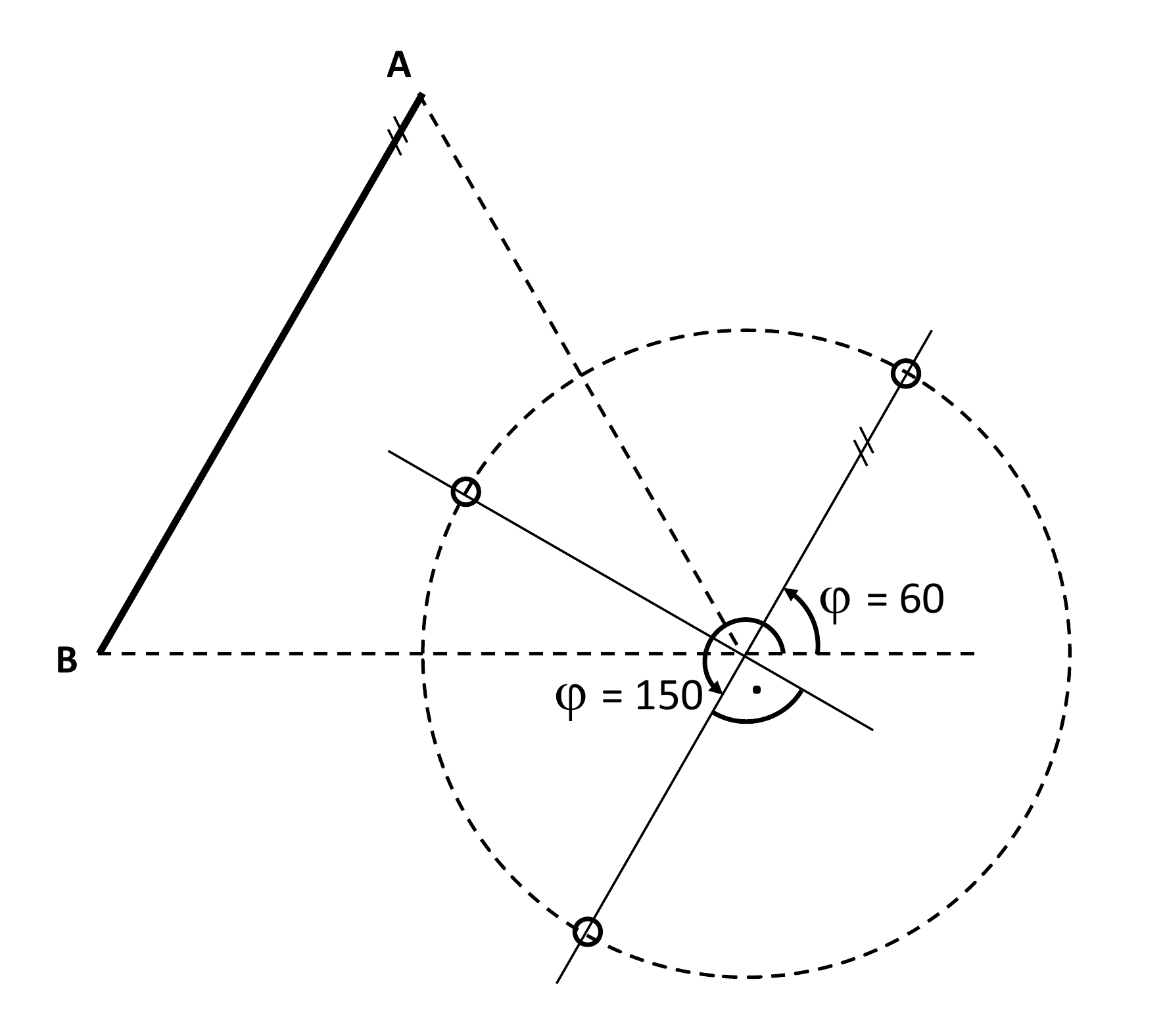}
 \caption{Left: Perturbations of the original equilateral triangle. With A and B fixed, the vertex C moves along a circle. The position of C is represented by the angle $\varphi$. Right: Three positions given by $\varphi = 60^\circ, 150^\circ, 240^\circ$ are indicated by small circles. Values $60^\circ$ and $150^\circ$ separate the region where the initial triangle has larger area than the equilateral triangle from the region with the opposite relation between the areas.} \label{fig:initial_deformations}
\end{figure}

The test is performed for different initial deformations of the original triangle. In Figure \ref{fig:initial_deformations} we show the initial triangles. Two vertices A and B are fixed. We place the third vertex C at various locations along a circle with the centre in the third vertex of original triangle and with radius 0.5. This way the resulting initial triangles cover acute as well as obtuse triangles. The position of C is determined by angle $\varphi$ ranging from $0^\circ$ to $360^\circ$.

\begin{figure}
\begin{center}
 \includegraphics[width=0.75\textwidth]{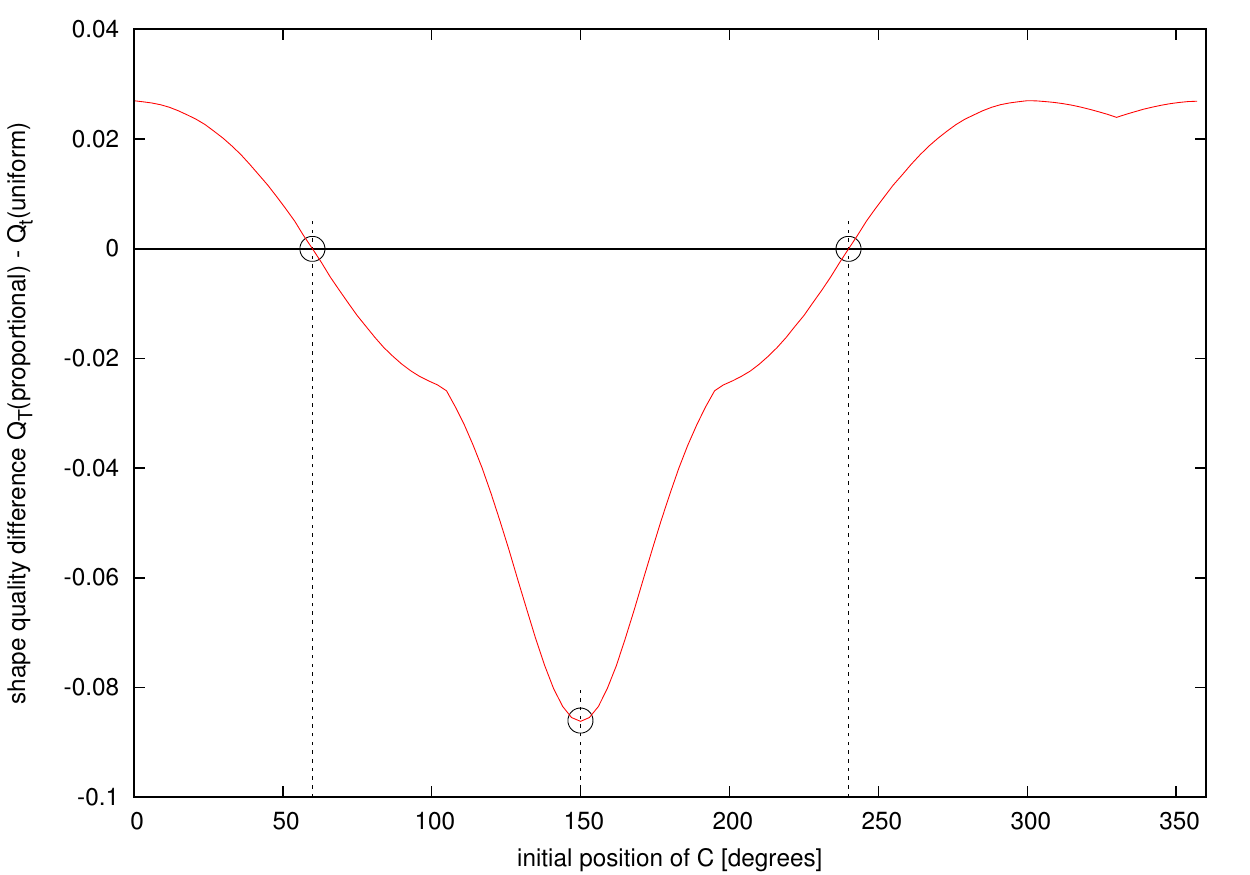}
 \caption{Difference between the quality of equilibrium triangles obtained by the new and original implementation. Small circles in the figure indicate three significant positions for C.} \label{fig:results}
 \end{center}
\end{figure}

From Figure \ref{fig:results} we clearly see when the proportional implementation performs better in terms of $\Qt$. At $0^\circ < \varphi < 60^\circ$ the proportional implementation results in better shape quality $\Qt$. Note that in this range, the initial deformed triangle has larger area than the original one. In the range $60^\circ < \varphi < 240^\circ$, the uniform implementation performs better with culmination point $\varphi = 150^\circ$. In this range, the initial deformed triangle has smaller area than the original one. In the range $\varphi = 240^\circ$ to $\varphi = 360^\circ$, the proportional implementation takes over. Here again, the surface of the initial triangle is larger than that of the original one.

In cases when the area of deformed triangle is larger than area of original triangle, proportional implementation gives better results, and otherwise the uniform implementation performs better. 

Let us first explain the latter statement. Suppose that the area of deformed triangle is smaller than that of the original one. In this case, the deformed triangle is stretched towards the original one. This situation is depicted in Figure \ref{fig:triangle_sim}. In the figure on the left, the stretched triangle $A'B'C'$ is similar to $ABC$, the angles do not change and thus the shape quality measure $\Qt$ does not change. On the contrary, with uniform implementation $\Qt$ changes. In the figure on the right, the lengths $|AA'|, |BB'|, |CC'|$ are equal, causing $|\angle C'A'B'| > |\angle CAB|$, which means that the updated angle is "less acute" than the one from the previous step. This leads to $\Qt' > \Qt$ and thus the result using uniform implementation resembles the equilateral original shape better.

The case when the deformed triangle has larger area than the original one is analogous. The deformed triangle is being shrunk, the proportional implementation keeps $\Qt$ constant, while the uniform implementation causes $|\angle C'A'B'| < |\angle CAB|$, i.e. the acute triangle becomes even "more acute", resulting in $\Qt' < \Qt.$ Therefore, in this case, the proportional implementation performs better.  

%%%%%%%%%%%%%%%%%%%%%%%%%%%%%%%%%%%%%%%
% 
%        section
%
%%%%%%%%%%%%%%%%%%%%%%%%%%%%%%%%%%%%%%%

\section{Simulation examples using only local area modulus}\label{sec:ellipsoids}
To demonstrate the effects of the proportional implementation of local area force, we show a 3D simulation experiment performed using our object-in-fluid module \cite{Cimrak2014} that is part of an open-source simulation package ESPResSo \cite{Arnold2013}. Visualizations were created using open-source analysis and visualization application ParaView \cite{Paraview2007}. In this experiment, we have an elastic object immersed in a stationary fluid that acts as a damping medium.

The relaxed shape of the object is a unitary sphere, because for a sphere we can check the return to the original shape. Out of the elastic moduli described in Section \ref{sec:forces}, only local area force is used in this comparison. The sphere is deformed to several initial shapes - ellipsoids, denoted by three numbers, e.g. $2 \times 1 \times 0.5$. This means that the $x$-semiaxis is stretched 2 times, $y$-semiaxis remains unchanged and $z$-semiaxis is halved. We also include isotropic inflation denoted by $1.5 \times 1.5 \times 1.5$. The deformed object is then left to relax using the uniform and proportional local area force. 

We have used a triangulation with 642 nodes and 1280 triangles with the average initial $Q_{\bar{T}0} \approx 0.93$. 

In these simulations, we have tracked the average final $Q_{\bar{T}}$ of all triangles and average distance of nodes to sphere surface. Representative results are summarised in Table \ref{tab:ellipsoids}. We do not include the shrinking deformations, such as ellipsoid $1 \times 0.75 \times 0.5$, because the relaxation after this kind of deformation results in "wrinkled" ellipsoids (using both implementations) as the expanding objects are unable to keep "flat" surface.  

\begin{table}
\caption{Characteristics of simulations with 5 different initial deformations, each run with both implementations of local area force: uniform and  proportional.}\label{tab:ellipsoids}
\centering
    \begin{tabular}{ c | r   r | r   r}
Deformation & \multicolumn{2}{c}{Average final $Q_{\bar{T}}$}  &  \multicolumn{2}{c}{Average dist to sphere}  \\  \hline
& uniform & proportional & uniform & proportional  \\ \hline
$3	\times 2		\times 1$ &	0.9266 &	0.9329 &	0.0269 &	0.0140 \\
$2	\times 1.5		\times 1$ &	0.9253 &	0.9324 &	0.0302 &	0.0162 \\
$2	\times 1		\times 0.5$ &	0.7492 &	0.7946 &	0.1560 &	0.1315 \\
$1.1	\times 1		\times 0.9$ &	0.9221 &  0.9222 &	0.0325 &	0.0325\\
$1.5	\times 1.5		\times 1.5$ &	0.9320 &  0.9341 &	0.0078 &  0.0079\\ \hline

    \end{tabular} 
\end{table}

\begin{figure}
 \includegraphics[width=\textwidth]{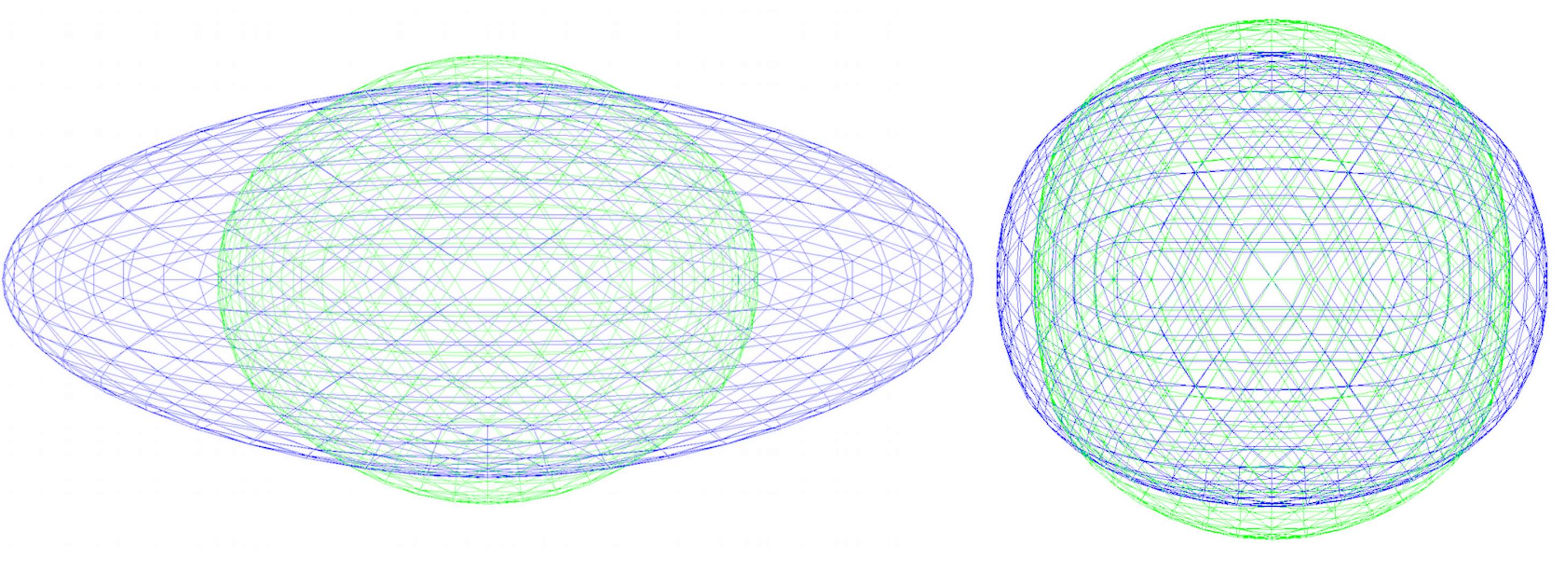}
 \caption{Relaxation of ellipsoid $3 \times 2 \times 1$, left: intermediate step, right: final shapes after particles stopped moving. The uniform implementation is represented in blue and  proportional in green.} \label{fig:ellipsoids}
\end{figure}

We see that the final $Q_{\bar{T}}$ is larger for the proportional implementation. 
Also, in all cases but the very last, the return to spherical shape is better or equal using the proportional implementation. Even in the case, where the uniform implementation gave better result, the proportional implementation was almost as good. The final restored spherical shape was measured as the average absolute value of node distances from spherical surface (see also Figure \ref{fig:ellipsoids}). 

Overall, these experiments have confirmed that the proportional implementation performs better not only for individual triangles, as shown in the analytical example in the previous section, but also for 3D objects. 

\section{Simulation example including other moduli}
\begin{figure}
 \includegraphics[width=\textwidth]{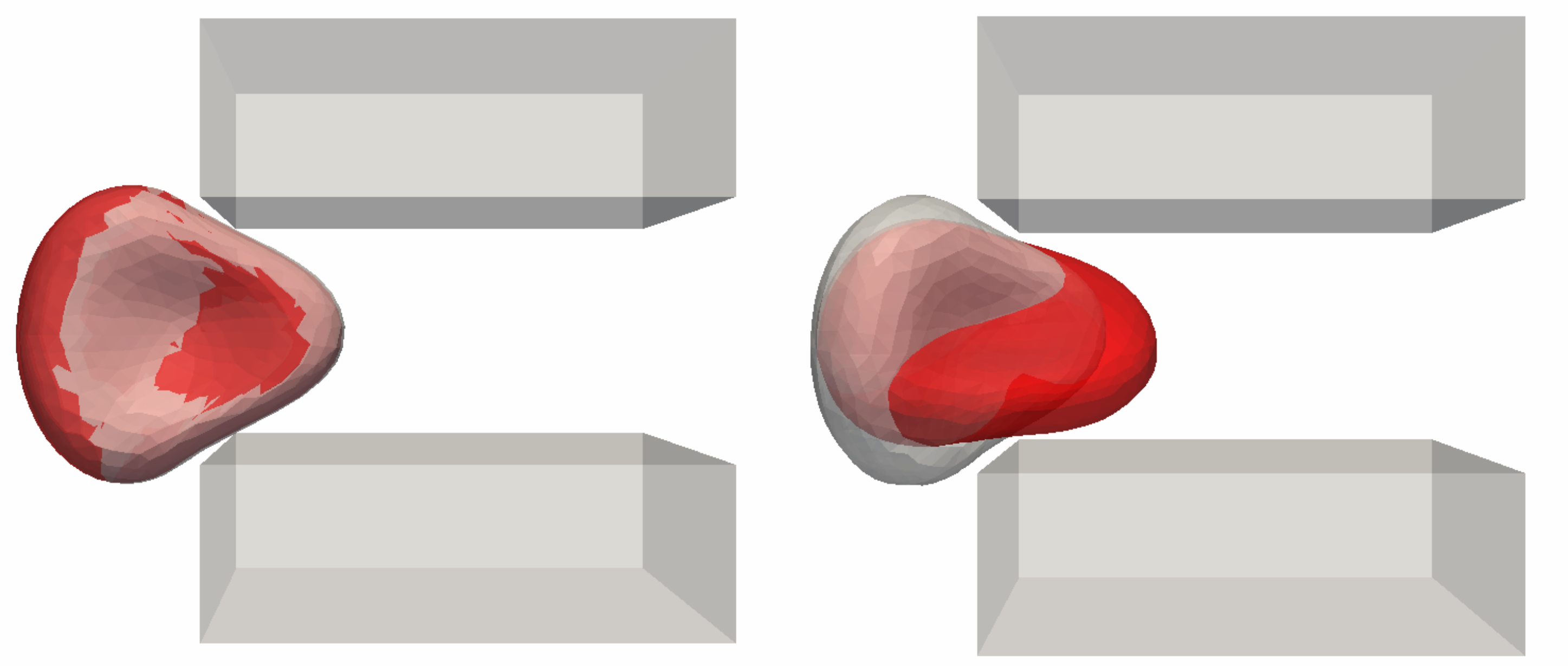}
 \caption{A red blood cell in flow entering a narrow opening in the channel. The uniform implementation is represented in grey, proportional in red. The shape of the cell as it comes to the narrow opening is the same in both implementations (left), however, when the fluid forces it to enter the opening, it deforms differently when using the two different implementations of local area (right).} \label{fig:narrow}
\end{figure}

While the simulations described in the previous section allow us to directly look at the effect of the various local area forces, the initial deformations are artificial. They would not be induced in simulations of actual immersed objects that conserve objects' global surface, volume and include shape-restoring bending forces. Therefore, we have also looked at simulations of a red blood cell in flow in a channel passing through a narrow opening, see Figure \ref{fig:narrow}, that induces "natural" deformations. 

The cell triangulation had 670 nodes, 1336 triangles and average initial $Q_{\bar{T}0} \approx 0.85$. The cell diameter was $7.82 \mu m$ and the channel dimensions were $70\times16\times16 \mu m^3$. The height of the narrow opening was $7 \mu m$. We have used the following values of elastic parameters: $k_s= 0.0045, k_b =0.155, k_{al} =0.003, k_{ag} =0.423$ and $k_v=1.25$ that were obtained by calibration using simulations of the optical tweezers experiment \cite{Tothova2015b}. The movement of fluid was induced by applying force density $f = 10^6 N/m^3$.  

We have observed that the shape of the cell as it comes to the narrow opening in the channel is the same in both implementations, however, when subjected to stress (fluid forcing it to enter the opening), it deforms differently when using the two different implementations of local area while keeping all other moduli the same, Figure \ref{fig:narrow}. So while the overall cell behavior is consistent, the definition of local area forces does have an influence on the behavior of the simulated cell.

%%%%%%%%%%%%%%%%%%%%%%%%%%%%%%%%%%%%%%%
% 
%        section
%
%%%%%%%%%%%%%%%%%%%%%%%%%%%%%%%%%%%%%%%

\section{Extension to global area and volume conservation}
The approach described in Section \ref{sec:definition} can also be used for global area force. We can define the proportional global area forces for one triangle using formula:
\begin{equation}\label{eq:new_global_area_equation}
F_{ag}(A)=\frac{t_A'}{t_A'^2+t_B'^2+t_C'^2}k_{ag}\Delta S_g
\end{equation}
and analogous for vertices B and C. The global area elastic coefficient has units $\left[\frac{N}{m}\right]$ and the corresponding elastic energy is shape independent. The action of this force is almost identical to the local area force with the only difference being that it is proportional to the difference of whole object's surface area form the relaxed surface area.

The situation is slightly different for volume conservation. Here the elastic coefficient $k_v$ corresponds to bulk modulus of the object that measures the object's resistance to uniform compression and thus has units \(\left[\frac{N}{m^2}\right]\). The elastic force pointing from vertex $A_i$ towards the centroid of the object derived using the procedure described in Section \ref{sec:definition} is:
\begin{equation}\label{eq:new_volume_equation}
F_{v}(A_i)=\frac{t_i'}{\sum_{j=1}^N t_j'^2}k_{v}\Delta V
\end{equation}
where $t_{i}'=|A_i'T|$, $i=1..N$, $T$ is the centroid of the object and $N$ the number of mesh nodes. 

%%%%%%%%%%%%%%%%%%%%%%%%%%%%%%%%%%%%%%%
% 
%        section
%
%%%%%%%%%%%%%%%%%%%%%%%%%%%%%%%%%%%%%%%

\section{Conclusion}
In this article, we have proposed a new approach to defining the elastic forces that conserve area. The main idea is an unequal distribution of acting force among the nodes. The non-uniformity accounts for the fact that some nodes are further from their preferred position than others and thus receive larger proportion of the acting force. Using an analytical example, we have shown that the proportional local area force offers improvement when the current area of triangle is greater than the relaxed area. Using simulations, we have demonstrated that this new implementation shows quantifiable and measurable improvement for large deformations of sphere and is consistent with the behavior of the previously known model of red blood cell subjected to deformation.

\end{document}